\documentclass[preprint,showpacs,preprintnumbers,amsmath,mprb]{revtex4-1}

\usepackage{graphicx}                
\usepackage{dcolumn}                 
\usepackage{bm}                      
\usepackage{hyperref}
\usepackage{color}
\usepackage{amsfonts}
\usepackage{upgreek,braket}
\usepackage{amsbsy}

\begin{document}

\title{ Phonon-Assisted Incoherent Excitation of a Quantum Dot and its Emission Properties}

\author{S. Weiler$^{\ast}$, A. Ulhaq, S.~M.~Ulrich, D.~Richter, M. Jetter and
P. Michler}

\affiliation{Institut f\"ur Halbleiteroptik und Funktionelle
Grenzfl\"achen, Universit\"at Stuttgart, Allmandring 3, 70569
Stuttgart, Germany,}

\author{C. Roy and S. Hughes}
\affiliation{Department of Physics, Engineering Physics and
Astronomy, Queens's University, Kingston, Ontario, Canada K7L 3N6}

$^{\star}$~Corresponding author: s.weiler@ihfg.uni-stuttgart.de;
Homepage: http://www.ihfg.uni-stuttgart.de


\begin{abstract}

We present a detailed study of a phonon-assisted incoherent
excitation mechanism of single quantum dots. A spectrally-detuned
laser couples to a quantum dot  transition by mediation of acoustic phonons,
whereby excitation efficiencies up to 20~\% with respect to
strictly resonant excitation can be achieved at T = 9~K. Laser
frequency-dependent analysis of the quantum dot intensity distinctly maps
the underlying  acoustic phonon bath and shows good agreement
with our polaron master equation theory. An analytical solution for the photoluminescence
  is introduced which  predicts a broadband incoherent coupling process  when electron-phonon scattering is in the  strong phonon coupling (polaronic) regime. Additionally, we investigate the coherence properties
of the emitted light and study the impact of the relevant pump and
phonon bath parameters.

\end{abstract}



\maketitle

Important properties of non-classical light emission from a single
quantum dot (QD), e.g. its exciton linewidth and coherence
properties, depend on the physical nature of the excitation process
of the QD. A standard method to optically address QDs is through
excitation into the barrier or wetting layer, which causes
subsequent capture and relaxation, finally resulting in
the recombination of carriers and the emission of photons. Such an
{\em incoherent pumping} mechanism leads to homogeneous broadening
of the excited state and results in a reduction of coherence time.
A more selective excitation of a single QD is possible via pumping
into a higher electronic (d-, p-shell) or the QD ground state
(s-shell). The
latter technique is suitable to generate close to Fourier-transform-limited photons \cite{Ates.Ulrich:2009}.\\

Due to distinct coupling of QD confined-state dynamics to the
surrounding solid-state crystal, phonon-mediated excitation offers
an alternative way of selective emitter state preparation. For
instance, pumping of single QDs via the energetically sharp
longitudinal-optical phonon (LO) resonance $\sim 35$~meV above the
dot ground state has been demonstrated
\cite{Lemaitre.Ashmore:2001,Toda.Moriwaki:2001,Findeis.Zrenner:2000,Outlon.Finley:2003}.
Besides LO coupling, other optical excitation methods rely on the
interaction of the electron-hole pair with acoustic phonons, for
instance non-resonant coupling (NRC) between emitter and
cavity~\cite{Hughes:PRB2011,Calic:PRL2011,Vuckovic:2011}. This
effect causes a detuned cavity mode to be efficiently excited by a
QD coupled to the surrounding acoustic phonon bath. The inverse
NRC effect, where the QD is excited via the cavity photon
emission, has also been demonstrated
experimentally~\cite{Kaniber.Neumann:2009}. Recent theoretical
analysis in the context of the NRC effect, however, have shown
that simple Lorentzian like pure-dephasing models are not
sufficient to fully explain this
phenomenon~\cite{Hohenester:2010,Hughes:PRB2011,Calic:PRL2011}.
Especially in the domain of resonance fluorescence, where the
QD-cavity system is excited coherently, significant phonon-induced
coupling between the QD exciton and the cavity is
predicted~\cite{Roy.Hughes:2011a}, resulting in phonon-mediated
excitation-induced dephasing (EID) and pronounced exciton-cavity
feeding. The former EID mechanism has been observed  in
micropillar-QD systems~\cite{stuttgart_prl}, and related EID
phenomena have been measured using coherently excited  QDs with
pulsed lasers~\cite{Ramsey:PRL2010a}; EID, via resonance
fluorescence,
 manifests in spectral broadening of the Mollow triplet sidebands as the strength of the pump field is increased;
 however, it does not give any direct information about the spectral characteristics of the broad phonon bath.
The exciton-acoustic phonon coupling is also directly observable
via the broad phonon bands
on the higher and lower energetic side of the zero-phonon line
(ZPL) that can be theoretically explained by the independent boson
model \cite{Mahan} under consideration of pure dephasing effects
\cite{Takagahara:1999}. The ZPL is Fourier-limited up to a first
approximation but higher-order coupling terms lead to a broadening
\cite{Muljarov.Zimmermann:2004}, which increases with temperature.
These phonon-based pure dephasing effects have been experimentally
studied in detail, particularly using the highly phase-sensitive
technique of four-wave mixing \cite{Borri.Langbein:2009}. The LA
phonon sidebands have also been directly observed in QD emission
spectra at elevated temperatures
\cite{Besombes.Khung:2001,Favero.Cassabois:2007}, using incoherent
excitation (i.e., pump laser excitation that is spectrally far
detuned from the target exciton state).\\

In this Letter, we present a joint experimental-theory
investigation of phonon-mediated incoherent excitation in the
polaron regime. The clear signatures of phonon-mediated excitation
is quite distinct to all previous attempts at exploring
electron-phonon interactions in QDs, and we present an unequivocal
and more direct probe of the phonon bath. We introduce an
analytical  model for the excited QD--phonon-bath system in a
planar sample such that the coupling between laser photons and a
QD is mediated by acoustic phonons in the framework of an
effective phonon master equation (ME)~\cite{Roy.Hughes:2011},
which is derived from a full polaron ME ~\cite{Roy.Hughes:2011a}.
We show that the effective QD intensity is a direct result of
phonon-mediated coupling which depends on the phonon density of
states at the laser excitation energy and the pump intensity of
the field. Moreover, this results in an incoherent excitation
process that is a direct signature of exciton-phonon coupling
effects beyond a weak exciton-phonon coupling approach---where
such a mechanism is absent~\cite{nazirPRB,KaerPRL}. Inspired by
related theoretical predictions~\cite{Ahn:PRB05,Roy.Hughes:2011},
we experimentally demonstrate the effective pumping of a single QD
via LA phonon coupling by spectrally tuning the laser close to the
QD s-shell resonance within the range of LA phonon energies. The
detuning-dependent frequency scans yield QD intensity profiles
exhibiting both the ZPL profile along with the broad LA phonon
sidebands (spanning more than 5 meV). Additionally we compare the coherence properties of the
QD emission, excited via this incoherent excitation process driven
by a coherent laser into the acoustic phonon bath
with that of a resonantly excited QD.\\

For our experiments, we employed self-assembled In(Ga,As)/GaAs QDs
grown by metal-organic vapor-phase epitaxy. A single layer of QDs
was centered in a 1$\lambda$-thick planar GaAs cavity surrounded
by alternating $\lambda/4$ periods of AlAs/GaAs as 4 top and 20
bottom distributed Bragg reflectors (DBR). The sample was mounted
in a He-flow cryostat at controllable temperatures of $T = (4 -
15)$~K. For resonant laser stray-light suppression a orthogonal
excitation-detection geometry micro photoluminescence ($\mu$PL)
setup in combination with a pinhole assembly and polarizer was
used. The sample was excited by a narrow band ($\sim$ 500~kHz)
tunable Ti:Sapphire continuous wave laser. More details of the
experimental techniques and the setup are given in
Refs.~\cite{Ates.Ulhaq:2009,
Ulhaq.Ates:2010,Ates.Ulrich:2009}.\\


Figure 1 illustrates the effect of phonon-assisted incoherent
excitation  of a QD. For the experimental conditions shown, the
laser excitation is energetically blue-detuned from the emitter by
$\Delta = \hbar \cdot (\omega_{L}-\omega_{\rm QD}) = 596$~$\mu$eV.
Even though the laser is not resonant with any higher QD shells
(energetic separation between s- and p-shell separation is
typically $\sim 25$~meV), considerable emitter intensity can be
observed in the $\mu$PL spectrum. The inset of Fig.~1 shows the
corresponding auto-correlation measurement of QD photons
unambiguously proving the single-photon emission nature with a
$g^{(2)}(0)$-value of $0.16 \pm 0.02$, deconvoluted with respect
to the setup time resolution of $450$~ps (convoluted value:
$g^{(2)}(0) = 0.35 \pm 0.04$).\\


To gain more insight into the effect of the phonon-induced
incoherent excitation, we scanned the cw laser over the QD resonance
in steps of $\sim15$~$\mu$eV. Under constant excitation power, a
$\mu$PL spectrum is taken at each step ($P = 500$~$\mu$W, $T =
5$~K). The PL data depicted in Fig.~2 reveal appreciable QD
emission over a long range of frequencies even away from the
s-shell resonance. The persistent presence of the QD signal for a
relatively large range of laser energies rules out excitation of
the QD via the quantized energy eigenstates of the dot itself.
Instead we attribute the excitation of the QD over a continuum of
frequencies to the presence of the LA phonon bath. The QD is
effectively excited by either emitting ($\Delta > 0$) or absorbing
($\Delta < 0$) phonons which compensate for the energy difference
between the excitation source and the exciton s-shell energy. The
inset of Fig.~2 (which shows a zoom into the region $\Delta
\approx 0$) displays the fine frequency scan over the QD s-shell
with a step size of $\sim2$~$\mu$eV. The increase of the signal
towards $\Delta = 0$ reflects the clear onset of resonance
fluorescence (RF) which overtakes the laser stray-light intensity
in the composite fluorescence and laser signal. Remarkably, such a
pronounced phonon-coupling effect occurs even without any cavity
coupling.\\


The effect of phonon-induced excitation is  modeled using a
polaronic  ME where explicit phonon-mediated
processes have been considered and derived in the form of
Liouvillian superoperators.
 A polaron ME has also been
used to investigate QD Rabi
oscillations~\cite{nazir2}.
 The effective phonon model is
explained in detail in Ref.~\cite{Roy.Hughes:2011}, which also
includes a cavity system. For the current system of interest
(i.e., with no cavity coupling),
 the  ME
is \cite{Roy.Hughes:2011}
\begin{eqnarray}
\frac{\partial \rho}{\partial t}& & =\frac{1}{i\hbar}[H_{S}',\rho(t)]+ \frac{\gamma}{2}{\cal L}[\sigma^-] +\frac{\gamma'}{2}{\cal L}[\sigma_{11}]+\frac{\Gamma_{\rm ph}^{\sigma^+}}{2}{\cal L}[\sigma^+] +\frac{\Gamma_{\rm ph}^{\sigma^-}}{2}{\cal L}[\sigma^-] ,
\label{effectiveME}
\end{eqnarray}
with a polaron-transformed system Hamiltonian, $H^{\prime}_{S} =
\hbar(-\Delta-\Delta_{P}){\sigma}^{+}{\sigma}^{-} + \hbar
\eta_{x}\langle B\rangle({\sigma}^{-}+{\sigma}^{+}) $, where
$\langle B\rangle=\exp\left [
-\frac{1}{2}\int^{\infty}_{0}d\omega\frac{J(\omega)}{\omega^{2}}\coth(\beta\hbar\omega/2)
\right ]$ is the thermally-averaged bath displacement operator and
$J(w)=\alpha_{p}\,\omega^{3}\exp
(-\frac{\omega^{2}}{2\omega_{b}^{2}})$ is the the characteristic
phonon spectral function; $\eta_x = 2 \Omega$ is the coherent pump
rate of the QD exciton,
$\sigma_{11}=\sigma^+\sigma^-$ with $\sigma^+,\sigma^-$ the Pauli operators, and
 we will incorporated the
polaron shift ($\Delta_P$) into the definition of $\omega_{QD}$ (and thus $\Delta$) below. The Lindblad operators,
${\cal L}[O]=2O\rho O^\dagger -O^\dagger O\rho - \rho O^\dagger O$,  describe dissipation through
ZPL radiative decay  and
ZPL pure dephasing ($\gamma'$), as well as pump-driven incoherent scattering processes mediated by the phonon bath:
%
%
$
\Gamma^{\sigma^+ / \sigma^-} _{\rm ph}= 2 \langle
B \rangle^2 \eta_x^2 \,{\rm Re} \left [ \int_{0}^\infty d\tau e^{\pm
i\Delta\tau} (e^{\phi(\tau)}-1) \right ],
$
where  $\phi(\tau)=\int^{\infty}_{0}d\omega\frac{J(\omega)}{\omega^{2}}
\left [ \coth(\hbar\omega/2k_bT)\cos(\omega \tau)-i\sin(\omega \tau)\right]$.
 Physically, the $\Gamma^{\sigma^{-}}_{\rm ph}$ process corresponds to an enhanced
 radiative decay, while the $\Gamma^{\sigma^{+}}_{\rm ph}$ process
 represents an incoherent excitation process. If the laser
 pump is within the vicinity of the phonon  bath, then the QD exciton
 can be excited through phonon emission or absorption~\cite{Roy.Hughes:2011}.
 As well as having analytical phonon scattering rates in the polaronic regime, we can also use Eq.~(\ref{effectiveME})
 to derive an explicit expression for the steady-state exciton population  (see Supplementary Material),
%
\begin{equation}
  \bar N_x  =
\frac{1}{2}\left [
1+ \frac{\Gamma_{\rm ph}^{\sigma^+}-\Gamma_{\rm ph}^{\sigma^-}-\gamma}
{\Gamma_{\rm ph}^{\sigma^+}+\Gamma_{\rm ph}^{\sigma^-}+\gamma+
\frac{4 \eta_x^{2}\braket{B}^2\Gamma_{\rm pol}}{\Gamma_{\rm pol}^2+\Delta^2}}
\right ],
\label{eq:nx}
\end{equation}
where $\Gamma_{\rm pol}=\frac{1}{2}(\Gamma^{\sigma^+}_{\rm
ph}+\Gamma^{\sigma^-}_{\rm ph}+\gamma+\gamma')$. For the  planar
system of interest, the QD intensity  from the vertical decay
channel of the sample is simply $I_{QD} \propto \bar N_x$. Importantly,
Eq.~(\ref{eq:nx}) includes the detuning  and pump dependence of the
phonon-induced scattering rates. We stress again that the incoherent excitation process described through
${\cal L}[\sigma^+]$ [see inset to Fig.~(4) for a schematic picture of this process] does not appear in ME approaches that assume
weak exciton-phonon coupling.
 Thus, our PL lineshapes (measured and predicted) show clear evidence of polaronic behaviour via phonon-bath-mediated incoherent excitation.


To compare with the theoretical predictions of Eq.~(\ref{eq:nx}),
the experimentally derived near-resonance $\mu$PL scans have been
evaluated in terms of normalized QD intensity versus detuning. The
results of a scan as shown in Fig.~2 are displayed in Figs.~3(a)
and 3(b) [black data points] revealing strong resonance
fluorescence of the QD near $\Delta \approx 0$ and a less intense,
but distinctly broad QD emission with a corresponding detuning range between
$\Delta \approx -1.5$~meV and $+2$~meV. The above presented
theoretical model has been fitted to the experimental data [red
solid lines in Figs.~3(a) and 3(b)]. The depicted profiles
consists of two parts, i.e., a rather {\em sharp} ZPL (Lorentzian
profile) at the QD resonance and a broader phonon-assisted
excitation feature around the ZPL. In the latter case, a distinct
asymmetry for $\Delta > 0$ is clearly visible. This is a direct
and unambiguous signature of the unequal probabilities for LA
phonon absorption and emission at low temperatures. This
incoherent scattering process is quite different to the incoherent
feeding that would result from a fast inter-level decay process.
To help identify this process further, we also plot the
calculation when the $\Gamma^{\sigma^+}_{\rm ph}$ process is
turned off, which confirms that the laser-driven incoherent
excitation process is the dominant phonon scattering process.

For modeling the measurement data, the corresponding values for
temperature $T$, pump rate $\eta_x$, and radiative decay $\gamma$,
have been experimentally derived by independent measurements and
are therefore fairly accurate values within their experimental
error. As is shown in the inset graph, $\eta_x$ can be extracted
from the HRPL data. For the conditions in Fig.~3(a) the center to
sideband Rabi-splitting $\Omega$ is found to be $(16.7 \pm 0.7)$
$\mu$eV (= $4.04 \pm 0.17$~GHz), giving $\eta_x = \frac{2
\Omega}{2 \pi} = (5.32 \pm 0.23)$~$\mu$eV. The radiative lifetime
for most of the QDs in the sample is found to be rather similar
due to no preferential radiative enhancement of selective QDs by
Purcell-like effects. Independent time-correlated photon counting
measurements have revealed a typical radiative decay time of
(750 - 850)~ps which gives $\gamma \approx (0.77 - 0.88) $~$\mu$eV. The
coupling constant describing the interaction between the exciton
and the LA phonons via deformation potential $\alpha_p$, the pure
dephasing rate $\gamma'$ and the cut-off frequency $\omega_b$
(proportional to the inverse of the typical electronic
localization length in the QD) \cite{nazirPRB} are then derived via
fitting. It should be mentioned that in most of our measurements
we observed somewhat higher QD intensities than theoretically
expected for intermediate positive detunings (as visible in
Fig.~3(a) around $\Delta = 2$~meV and 3(b) around $\Delta =
1.7$~meV). This effect might be attributed to some additional
dephasing apart from the LA phonon coupling which could be caused
by a variety of possible effects like phonon scattering from
defects or trapped charges in the vicinity of the QD
\cite{Favero.Cassabois:2007}. Excitation via phonon-coupling is
observed to be rather  efficient and approximately 20\,\% of the
QD intensity is achieved with respect to strictly resonant
excitation with the distinct advantage that the laser stray-light
can be easily separated from the QD emission;  this
efficient phonon-induced excitation is
unique to the QD environment and is substantially different to exciting an atomic resonance.\\

Photon-visibility measurements have been performed by Michelson
interferometry to determine the coherence properties of photons
emitted by both resonant and off-resonant excitation conditions.
Figure 3(c) shows the sinusoidally-varying visibility of the QD
emission under strictly resonant excitation which indicates the
presence of the Mollow triplet \cite{Muller.Flagg:2007,Mollow} in
the spectral regime. The data has been fitted by the following
function $g^{(1)}(\tau)= \frac{1}{2}e^{-\Gamma_{\rm pol}\tau}
e^{i\omega\tau} + \frac{1}{4}e^{-(\Gamma_{\rm tot}/2)\tau}
e^{i(\omega-\Omega)\tau} +\frac{1}{4}e^{-(\Gamma_{\rm tot}/2)\tau}
e^{i(\omega+\Omega)\tau}$, with $\Gamma_{\rm tot} = \Gamma_{\rm pol}+
\Gamma_{\rm pop}$ and $\Gamma_{\rm pop}= \gamma + \Gamma^{\sigma^+}_{\rm
ph} + \Gamma^{\sigma^-}_{\rm ph}$ , where the exponential decay
reflects the limited coherence yielding a $T_2 = (258 \pm 31)$~ps.
Even under strictly resonant excitation the coherence is not
lifetime limited, which may be attributed to spectral diffusion
effects. As a comparison, we experimentally investigated the
emission coherence under the non-resonant excitation conditions of
phonon-assisted excitation for different detunings $\Delta$. The
Gaussian fitting functions applied to the measurement results
shown in Fig.~3(d-f) reveal coherence times of $T_2 \sim 20$~ps
rather limited compared to the value under resonant pump due to
dephasing induced by the excitation mechanism.\\


To gain further insight into the effect of phonon-assisted
incoherent excitation, we have systematically studied
theoretically the effects of $\omega_b$, $T$, $\alpha_p$, $\eta_x$
on the resulting intensity profiles in Figs.~4 (a-d). An increase
in the cut-off frequency $\omega_b$ (i.e., a decrease in QD size)
leads to a blue shift of the phonon reservoir replica of the QD
intensity profile.  In contrast, increasing temperature $T$, pump
rate $\eta_x$ or coupling factor $\alpha_p$ overall increases the
QD intensity for off-resonant excitation conditions. This can also
be seen in Figs.~3 (a) and (b), where increased temperature and
pump rate leads to a higher emission efficiency in 3(b) as
compared to 3(a). For increasing temperature, these features
become more symmetric due to increasing phonon state occupations.
Parameters $\alpha_p$ and $\eta_x$ have similar effects on the
shape of the intensity profile but keep the asymmetry unchanged.
Variation of $\gamma,\gamma'$ (not shown) mainly affects the width
of the ZPL and has almost negligible effect on the broader
intensity profile. We would like to emphasize that the effect has
been observed consistently for several QDs in the sample. The
investigations are also particularly important for experiments
under pulsed resonant excitation. Here the spectrally broad laser
(compared to cw excitation) has distinct overlap not only with the
ZPL, but also with the acoustic phonon side wings. Therefore the
resonance fluorescence signal has additional contributions from
photons emitted after phonon-induced coupling, which might
increase the excitation efficiency but also change the coherence
properties of
the detected photons.\\

In conclusion, we have presented a detailed study of the
phonon-assisted incoherent excitation effect for self-assembled
QDs. The experimentally investigated dot intensity as a function
of laser-QD detuning is in very good agreement with a polaronic ME
model.
%
Additionally, we have studied the coherence properties
of QD emission via the phonon-mediated excitation and the
influence of different realistic parameters on the spectral shape
of the intensity profile has been theoretically investigated,
using a newly presented analytical solution for the steady-state
exciton population. Phonon-assisted incoherent excitation therefore does
not only provide a unique excitation mechanism of a semiconductor
QD but also an effective new tool to map the characteristic
features of the phonon bath
present in such a solid state quantum-emitter system.
This process is also polaronic in nature, demonstrating exciton-phonon coupling effects beyond the weak coupling regime.\\

We thank W.-M. Schulz for expert sample preparation. S. Weiler
acknowledges financial support by the Carl-Zeiss-Stiftung. A.
Ulhaq acknowledges funding from International Max Planck Research
School IMPRS-AM.
This work was supported by the National Sciences and Engineering Research Council
 of Canada.


\begin{thebibliography}{31}
\expandafter\ifx\csname
natexlab\endcsname\relax\def\natexlab#1{#1}\fi
\expandafter\ifx\csname bibnamefont\endcsname\relax
  \def\bibnamefont#1{#1}\fi
\expandafter\ifx\csname bibfnamefont\endcsname\relax
  \def\bibfnamefont#1{#1}\fi
\expandafter\ifx\csname citenamefont\endcsname\relax
  \def\citenamefont#1{#1}\fi
\expandafter\ifx\csname url\endcsname\relax
  \def\url#1{\texttt{#1}}\fi
\expandafter\ifx\csname
urlprefix\endcsname\relax\def\urlprefix{URL }\fi
\providecommand{\bibinfo}[2]{#2}
\providecommand{\eprint}[2][]{\url{#2}}

\bibitem[{\citenamefont{\textit{et al.}}(2009)}]{Ates.Ulrich:2009}
  \bibinfo{author}{\bibfnamefont{S. Ates}, \bibfnamefont{S. M. Ulrich},
  \bibfnamefont{S. Reitzenstein}, \bibfnamefont{A. L\"{o}ffler},
  \bibfnamefont{A. Forchel}, and \bibfnamefont{P. Michler}},
  \bibinfo{journal}{Phys. Rev. Lett.} \textbf{\bibinfo{volume}{103}},
  \bibinfo{pages}{167402} (\bibinfo{year}{2009}).

\bibitem[{\citenamefont{\textit{et al.}}(2001)}]{Lemaitre.Ashmore:2001}
  \bibinfo{author}{\bibfnamefont{A. Lema\^{i}tre}, \bibfnamefont{A. D. Ashmore}, \bibnamefont{J. J. Finley}, \bibnamefont{D. J. Mowbray}, \bibnamefont{M. S. Skolnick}, \bibnamefont{M. Hopkinson}, and \bibnamefont{T. F. Krauss}},
  \bibinfo{journal}{Phys. Rev. B} \textbf{\bibinfo{volume}{63}},
  \bibinfo{pages}{161309(R)} (\bibinfo{year}{2001}).

\bibitem[{\citenamefont{\textit{et al.}}(1999)}]{Toda.Moriwaki:2001}
  \bibinfo{author}{\bibfnamefont{Y. Toda}, \bibfnamefont{O. Moriwaki}, \bibnamefont{M. Nishioka}, and \bibfnamefont{Y. Arakawa}},
  \bibinfo{journal}{Phys. Rev. Lett.} \textbf{\bibinfo{volume}{82}},
  \bibinfo{pages}{4114} (\bibinfo{year}{1999}).

  \bibitem[{\citenamefont{\textit{et al.}}(2000)}]{Findeis.Zrenner:2000}
  \bibinfo{author}{\bibfnamefont{F. Findeis}, \bibfnamefont{A. Zrenner}, \bibnamefont{G. B\"ohm}, and \bibfnamefont{G. Abstreiter}},
  \bibinfo{journal}{Phys. Rev. B} \textbf{\bibinfo{volume}{61}},
  \bibinfo{pages}{10579(R)} (\bibinfo{year}{2000}).

  \bibitem[{\citenamefont{\textit{et al.}}(2003)}]{Outlon.Finley:2003}
  \bibinfo{author}{\bibfnamefont{R. Oulton}, \bibfnamefont{J. J. Finley}, \bibnamefont{A. I. Tartakovskii}, \bibnamefont{D. J. Mowbray}, \bibnamefont{M. S. Skolnick}, \bibnamefont{M. Hopkinson},
  \bibnamefont{A. Vasanelli}, \bibnamefont{R. Ferreira}, and \bibfnamefont{G. Bastard}},
  \bibinfo{journal}{Phys. Rev. B} \textbf{\bibinfo{volume}{68}},
  \bibinfo{pages}{235301} (\bibinfo{year}{2003}).

\bibitem{Hughes:PRB2011}
S. Hughes, P. Yao, F. Milde, A. Knorr, D. Dalacu, K. Mnaymneh, V.
Sazonova, P. J. Poole, G. C. Aers, J. Lapointe, R. Cheriton, and
R. L. Williams, Phys. Rev. B {\bf 83}, 165313 (2011).

\bibitem{Calic:PRL2011}
M. Calic, P. Gallo, M. Felici, K. A. Atlasov, B. Dwir, A. Rudra, G. Biasiol, L. Sorba, G. Tarel, V. Savona, and E. Kapon,
Phys. Rev. Lett. {\bf 106}, 227402 (2011).


\bibitem[{\citenamefont{\textit{et al.}}(2011)}]{Vuckovic:2011}
  \bibinfo{author}{\bibfnamefont{A. Majumdar}, \bibfnamefont{E. D. Kim}, \bibfnamefont{Y. Gong}, \bibfnamefont{M. Bajcsy},
 and \bibfnamefont{J. Vu\v{c}kovi\'{c}}},
  \bibinfo{journal}{Phys. Rev. B} \textbf{\bibinfo{volume}{84}},
  \bibinfo{pages}{085309} (\bibinfo{year}{2011}).

\bibitem[{\citenamefont{\textit{et al.}}(2009)}]{Kaniber.Neumann:2009}
  \bibinfo{author}{\bibfnamefont{M. Kaniber}, \bibfnamefont{A. Neumann}, \bibfnamefont{A. Laucht}, \bibfnamefont{M. F. Huck}, \bibfnamefont{M. Bichler}, \bibfnamefont{M.-C. Amann} and \bibfnamefont{J. J. Finley}},
  \bibinfo{journal}{New Journal of Physics} \textbf{\bibinfo{volume}{11}},
  \bibinfo{pages}{013031} (\bibinfo{year}{2009}).

\bibitem[{\citenamefont{\textit{et al.}}(2010)}]{Hohenester:2010}
  \bibinfo{author}{\bibfnamefont{U. Hohenester}},
  \bibinfo{journal}{Phys. Rev. B}, {\bf \bibinfo{volume}{81}}, \bibinfo{pages}{155303} (\bibinfo{year}{2010}).



\bibitem{Roy.Hughes:2011a}
C. Roy and S. Hughes, Phys. Rev. Lett. {\bf 106}, 247403 (2011).


\bibitem{stuttgart_prl}S. M. Ulrich, S. Ates, S. Reitzenstein, A. L\"offler, A. Forchel and P. Michler,
Phys. Rev. Lett. {\bf 106}, 247402 (2011).



\bibitem{Ramsey:PRL2010a}
A. J. Ramsay, Achanta Venu Gopal, E. M. Gauger, A. Nazir, B. W.
Lovett, A. M. Fox, and M. S. Skolnick,
Phys. Rev. Lett. {\bf 104}, 017402 (2010).


\bibitem{Mahan}
  \bibinfo{author}{\bibfnamefont{G. D. Mahan}} \textit{Many particle physics}, (Springer, Berlin Heidelberg, 2000)


\bibitem[{\citenamefont{\textit{et al.}}(1999)}]{Takagahara:1999}
  \bibinfo{author}{\bibfnamefont{T. Takagahara}}, \bibinfo{journal}{Phys. Rev. B} \textbf{\bibinfo{volume}{60}},
  \bibinfo{pages}{2638} (\bibinfo{year}{1999}).


\bibitem[{\citenamefont{\textit{et al.}}(2004)}]{Muljarov.Zimmermann:2004}
  \bibinfo{author}{\bibfnamefont{E. A. Muljarov}, and \bibfnamefont{R. Zimmermann}},
  \bibinfo{journal}{Phys. Rev. Lett.} \textbf{\bibinfo{volume}{93}},
  \bibinfo{pages}{237401} (\bibinfo{year}{2004}).

\bibitem[{\citenamefont{\textit{et al.}}(2005)}]{Borri.Langbein:2009}
  \bibinfo{author}{\bibfnamefont{P. Borri}, \bibfnamefont{W. Langbein}, \bibfnamefont{U. Woggon}, \bibfnamefont{V. Stavarache}, \bibfnamefont{D. Reuter} and \bibfnamefont{A. D. Wieck}},
  \bibinfo{journal}{Phys. Rev. B} \textbf{\bibinfo{volume}{71}},
  \bibinfo{pages}{115328} (\bibinfo{year}{2005}).

\bibitem[{\citenamefont{\textit{et al.}}(2001)}]{Besombes.Khung:2001}
  \bibinfo{author}{\bibfnamefont{L. Besombes}, \bibfnamefont{K. Kheng},
  \bibfnamefont{L. Marsal}, and \bibfnamefont{H. Mariette}},
  \bibinfo{journal}{Phys. Rev. B} \textbf{\bibinfo{volume}{63}},
  \bibinfo{pages}{155307} (\bibinfo{year}{2001}).

\bibitem[{\citenamefont{\textit{et al.}}(2003)}]{Favero.Cassabois:2007}
  \bibinfo{author}{\bibfnamefont{I. Favero}, \bibnamefont{G. Cassabois}, \bibfnamefont{R. Ferreira},
  \bibfnamefont{D. Darson}, \bibfnamefont{C. Voisin}, \bibfnamefont{J. Tignon},
  \bibfnamefont{C. Delalande}, \bibfnamefont{G. Bastard},
  \bibfnamefont{Ph. Roussignol}, and \bibfnamefont{J. M. G\'{e}rard}},
  \bibinfo{journal}{Phys. Rev. B} \textbf{\bibinfo{volume}{68}},
  \bibinfo{pages}{233301} (\bibinfo{year}{2003}).

\bibitem[{\citenamefont{\textit{et al.}}(2011)}]{Roy.Hughes:2011}
  \bibinfo{author}{\bibfnamefont{C. Roy} and \bibfnamefont{S. Hughes}},
  \bibinfo{journal}{Phys. Rev. X} \textbf{\bibinfo{volume}{1}},
  \bibinfo{pages}{021009} (\bibinfo{year}{2011}).



\bibitem{nazirPRB}A. Nazir,
Phys. Rev. B {\bf 78}, 153309 (2008).

\bibitem{KaerPRL}
A. Moelbjerg, P. Kaer, M. Lorke, and J. M\o rk,
Phys. Rev. Lett. {\bf 108}, 017401 (2012)


\bibitem{Ahn:PRB05}
K. J. Ahn, J. F\"orstner, and A. Knorr,
Phys. Rev. B {\bf 71}, 153309 (2005).


\bibitem[{\citenamefont{\textit{et al.}}(2009)}]{Ates.Ulhaq:2009}
  \bibinfo{author}{\bibfnamefont{S. Ates}, \bibfnamefont{S. M. Ulrich},
  \bibfnamefont{A. Ulhaq}, \bibfnamefont{S. Reitzenstein},
  \bibfnamefont{A. L\"{o}ffler}, \bibfnamefont{S. H\"{o}fling},
  \bibfnamefont{A. Forchel}, and \bibfnamefont{P. Michler}},
  \bibinfo{journal}{Nature Phot.} \textbf{\bibinfo{volume}{3}},
  \bibinfo{pages}{724} (\bibinfo{year}{2009}).

\bibitem[{\citenamefont{\textit{et al.}}(2010)}]{Ulhaq.Ates:2010}
  \bibinfo{author}{\bibfnamefont{A. Ulhaq}, \bibfnamefont{S. Ates},
  \bibfnamefont{S. Weiler}, \bibfnamefont{S. M. Ulrich}, \bibfnamefont{S. Reitzenstein},
  \bibfnamefont{A. L\"offler}, \bibfnamefont{S. H\"ofling},
  \bibfnamefont{L. Worschech}, \bibfnamefont{A. Forchel}, and \bibfnamefont{P. Michler}},
  \bibinfo{journal}{Phys. Rev. B} \textbf{\bibinfo{volume}{82}},
  \bibinfo{pages}{045307} (\bibinfo{year}{2010}).

\bibitem{nazir2}D. P. S. McCutcheon and A. Nazir,
New J. Phys. {\bf 12}, 113042 (2010).

\bibitem[(1969)]{Mollow}
  \bibinfo{author}{\bibfnamefont{B. R. Mollow}},
  \bibinfo{journal}{Phys. Rev.}, \bibinfo{volume}{188} (\bibinfo{year}{1969}).


\bibitem[{\citenamefont{\textit{et al.}}(2007)}]{Muller.Flagg:2007}
  \bibinfo{author}{\bibfnamefont{A. Muller}, \bibfnamefont{E. B. Flagg}, \bibfnamefont{P. Bianucci},
  \bibfnamefont{X. Y. Wang}, \bibfnamefont{D. G. Deppe}, \bibfnamefont{W. Ma},
  \bibfnamefont{J. Zhang}, \bibfnamefont{G. J. Salamo}, \bibfnamefont{M. Xiao}, and \bibfnamefont{C. K. Shih}},
  \bibinfo{journal}{Phys. Rev. Lett.} \textbf{\bibinfo{volume}{99}},
  \bibinfo{pages}{187402} (\bibinfo{year}{2007}).






\end{thebibliography}


\newpage

\textbf{Figures}

\vspace{0.5cm}

\textbf{Figure~1, S. Weiler, P.R.L.}

\begin{figure}[!h]
\begin{center}
\includegraphics[width=12.0cm]{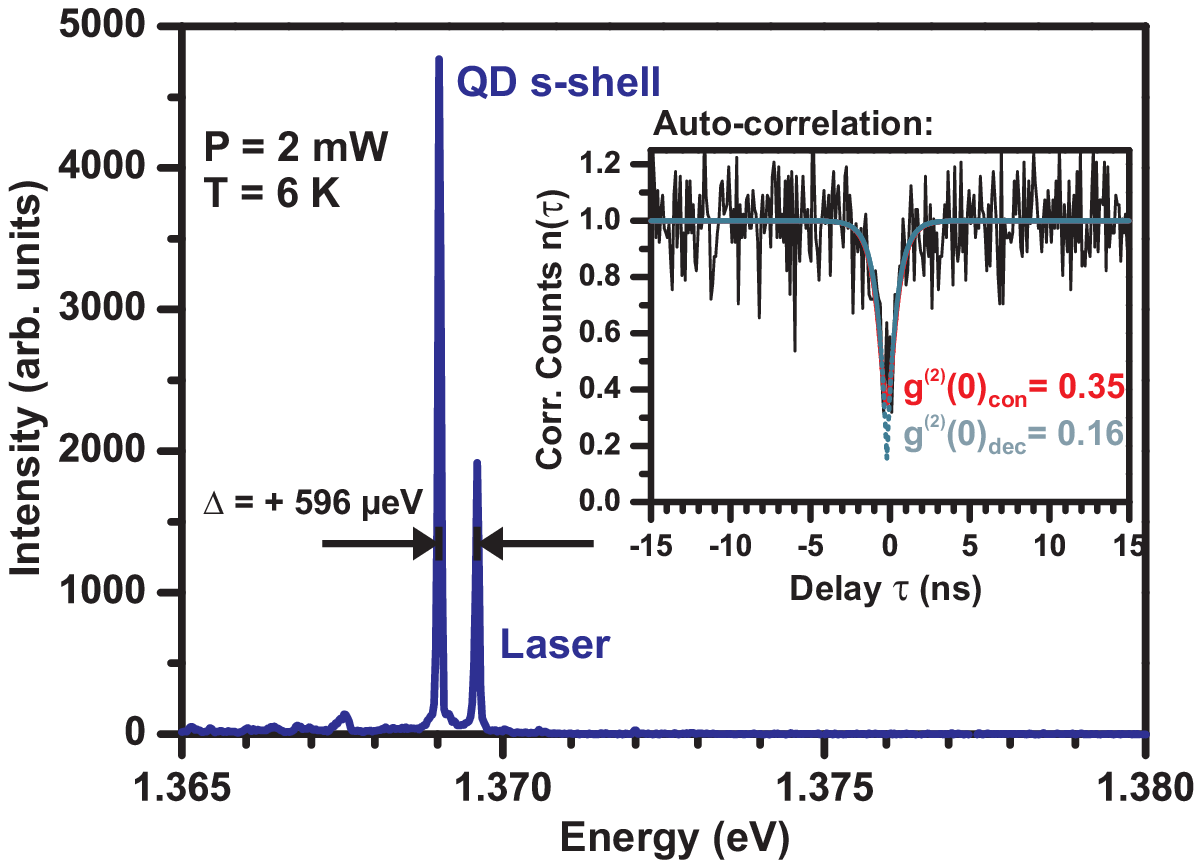}
\label{Figure1} \vspace{-0.3cm}
\end{center}
\end{figure}

\newpage
\textbf{Figure~2, S. Weiler, P.R.L.}

\begin{figure}[!h]
\label{Figure2}
\begin{center}
\includegraphics[width=12.0cm]{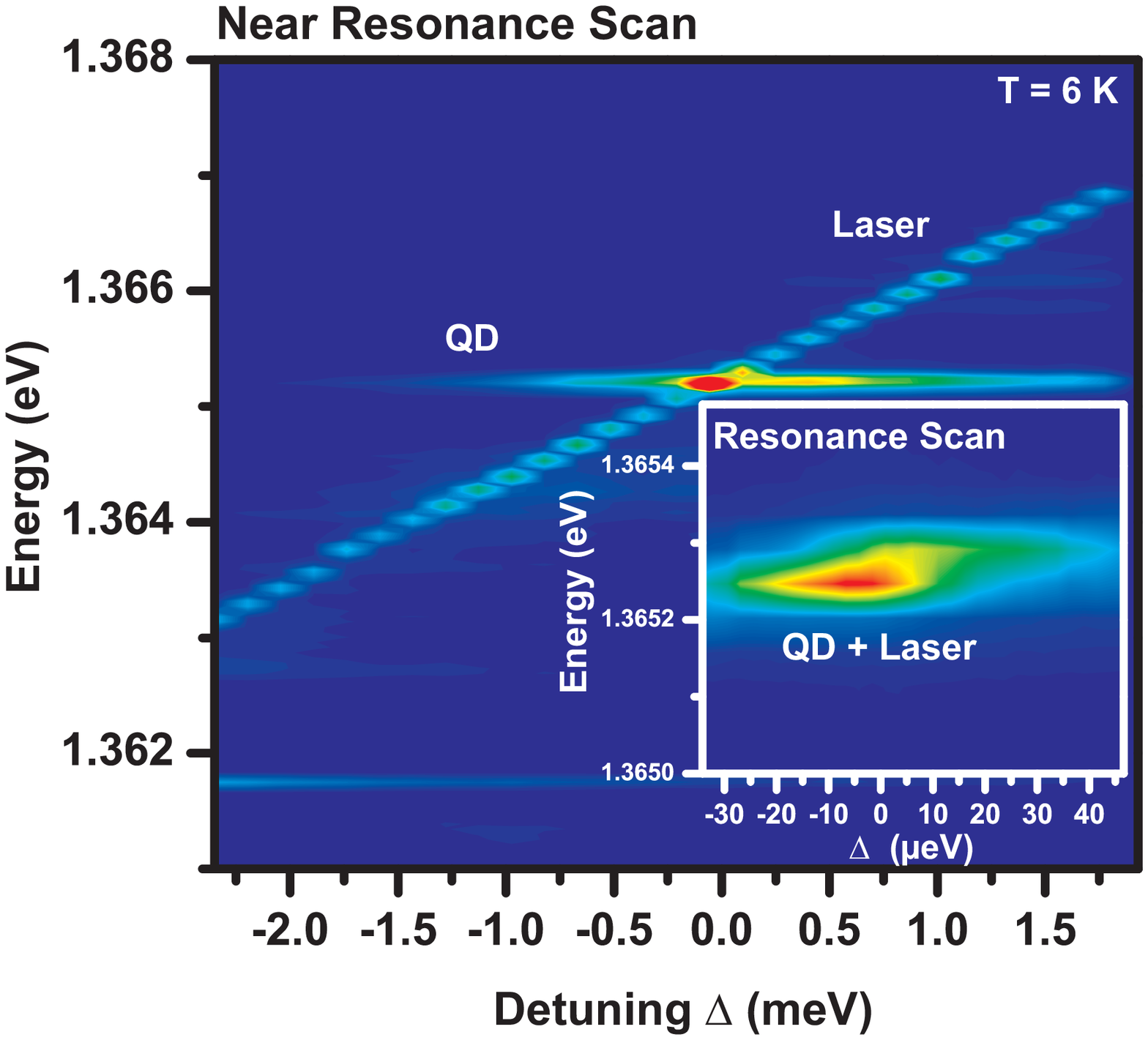}
\vspace{-0.3cm}
\end{center}
\end{figure}

\newpage
\textbf{Figure~3, S. Weiler, P.R.L.}

\begin{figure}[!h]
\label{Figure3}
\begin{center}
\includegraphics[width=12.0cm]{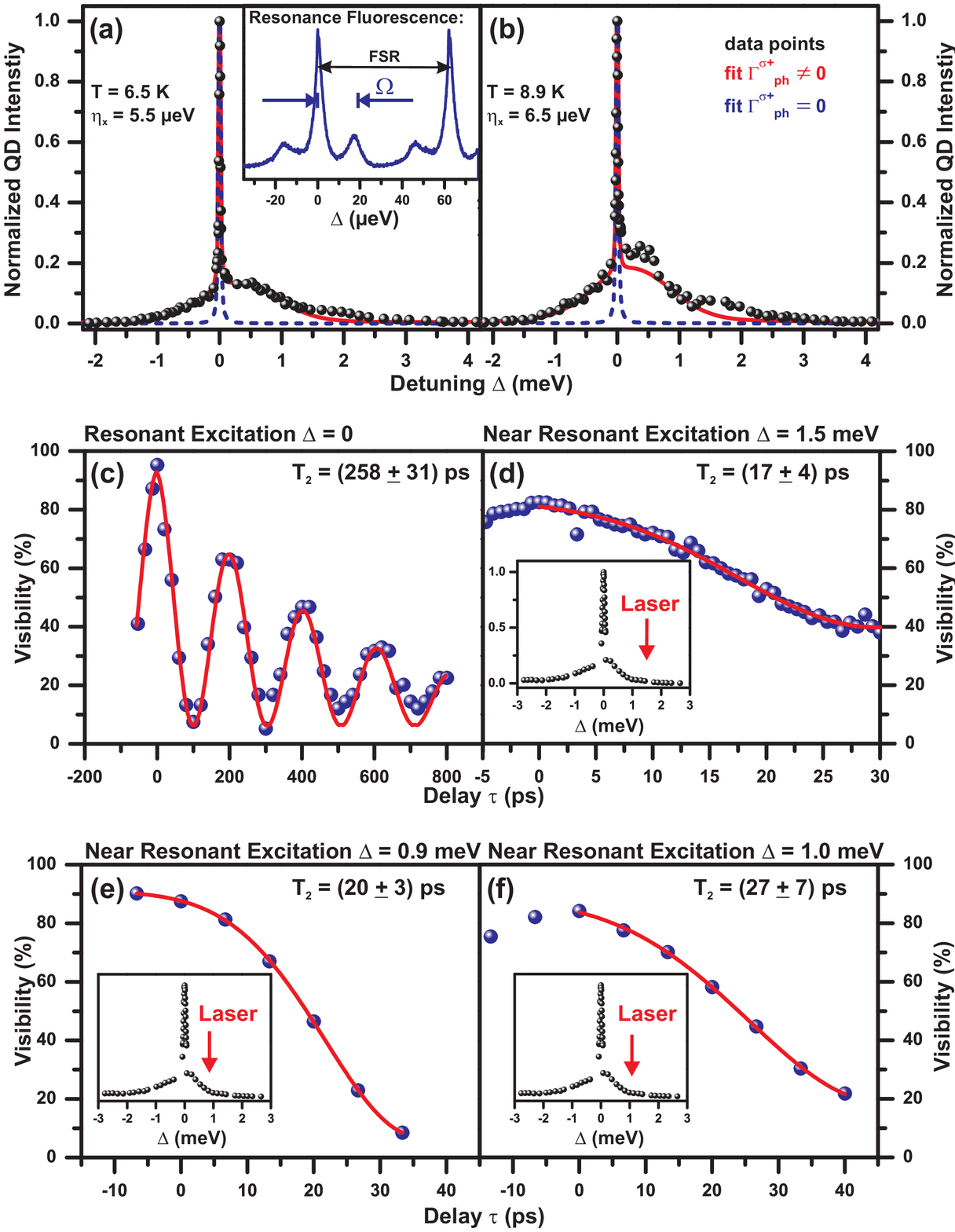}
\vspace{-0.3cm}
\end{center}
\end{figure}

\newpage
\textbf{Figure 4, S. Weiler, P.R.L.}

\begin{figure}[!h]
\begin{center}
\label{Figure4}
\includegraphics[width=12.0cm]{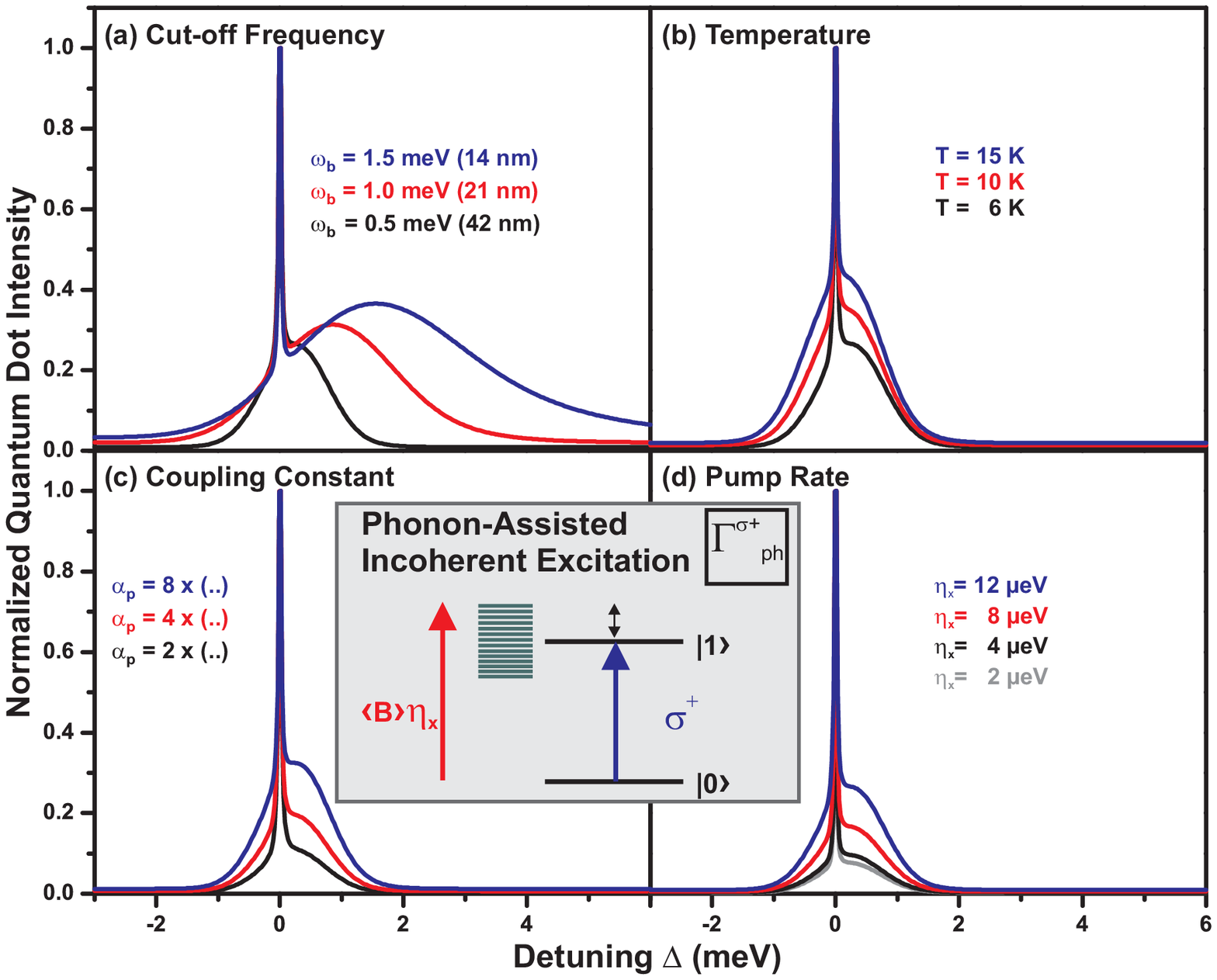}
\vspace{-0.3cm}
\end{center}
\end{figure}


\newpage
\textbf{Figure Captions} \vspace{0.5cm}

\textbf{FIG.1} Demonstration of phonon-assisted incoherent
excitation of a single QD. PL emission sprectrum of a QD
non-resonantly excited via a cw laser ($\Delta = 596$~meV)  Inset:
Corresponding auto-correlation measurement on the QD s-shell
proving almost backgroundfree single-photon emission with a
deconvoluted antibunching value (detector response) of
$g^{(2)}(0)_{\rm dec}=0.16 \pm 0.02$ (convoluted: $g^{(2)}(0)_{\rm
con}=0.35 \pm 0.04$).

\vspace{1cm}

\textbf{FIG.2} Near resonance frequency scan: The excitation laser
energy has been systematically varied in steps of $\sim
15$~$\mu$eV to scan over the QD s-shell at a fixed power of P =
$500$~$\mu$W. As evident from the color plot, emission from the QD
can be continuously traced. Inset:  Resonance scan: Detailed scan
over the QD s-shell in steps of $\sim 2$~$\mu$eV revealing the
onset of RF around $\Delta = 0$.

\vspace{1cm}

\textbf{FIG.3} (a)-(b) Intensity profiles: Integrated QD intensity
derived from a frequency scan (similar to that in Fig.~2) plotted
as a function of laser-QD detuning. The scans have been performed
at different temperatures. Experimental data indicated by black
circles and theory (red solid lines). The corresponding values
used to fit the data with our theory (with phonon-induced
processes $\Gamma^{\sigma^+}_{\rm ph}$ and $\Gamma^{\sigma^-}_{\rm
ph}$: solid (red) line, with only the process
$\Gamma^{\sigma^-}_{\rm ph}$: dashed (blue) line) are the cut-off
frequency $\omega_b = 0.6$~meV, coupling parameter $\alpha_p = 6
\times 0.06 \times \pi^2$~ps, rate of radiative decay rate $\gamma
= 0.82~ \mu$eV ($803$~ps) and pure dephasing rate $\gamma'=0.6~
\mu$eV ($1097$~ps). The thermally-averaged bath displacement
operator is calculated to be $\langle B \rangle = 0.91$ for the
conditions in (a) and $\langle B \rangle =0.87$ for (b). Inset:
HRPL spectrum with the characteristic RF spectrum in the frequency
domain, revealing a Rabi energy (center to sideband) of $\Omega =
(16.7 \pm 0.7)$~$\mu$eV. (c) Visibility measurements under
strictly resonant and off-resonant (d-f) phonon-assisted
incoherent excitation conditions. The off-resonant emission
reveals a distinctly shorter $T_2$ of ($17 \pm 4$)~ps for $\Delta
= 1.5$~meV, ($20 \pm 3$)~ps for $\Delta = 0.9$~meV and ($27 \pm
7$)~ps for $\Delta = 1.0$~meV, in comparison to the resonant
coherence of ($258 \pm 31$)~ps attributed to the phonon emission
dephasing process. The solid lines are the corresponding gaussian
fits to the experimental data.

\vspace{1cm}

\textbf{FIG.4} Investigation of the influence of the relevant
parameters on the intensity profile of the QD emission. While
keeping the other parameters fixed ($\omega_b = 0.5 $~meV, $T =
6$~K, $\gamma = 0.8$~$\mu$eV ($823$~ps), $\gamma' = 0.8$~$\mu$eV,
$\eta_x = 12$~$\mu$eV), the (a) phonon bath cut-off frequency
$\omega_b,$ (b) temperature $T$, (c) coupling factor $\alpha_p$,
and (d) pump rate $\eta_x = 2\,\Omega$ have been systematically
increased from bottom to top respectively. Inset: Illustration of
the incoherent excitation process (red arrow),
$\Gamma^{\sigma^+}_{\rm ph}$ scattering, mediated by the acoustic
phonon bath (green lines).

\end{document}